\documentclass[aps,physrev,preprint,groupedaddress]{revtex4-2}

\usepackage[english]{babel}
\usepackage{amsmath}
\usepackage{graphicx}
\usepackage{gensymb}
\usepackage{textcomp}
\usepackage[export]{adjustbox}
\usepackage{float}
\usepackage{siunitx}
\usepackage{csvsimple}
\usepackage{caption}
\usepackage{subcaption}
\usepackage{setspace}
\usepackage{pgffor}  
\usepackage[colorlinks=true, allcolors=blue]{hyperref}
\newcommand{\mytilde}{\raise.17ex\hbox{$\scriptstyle\mathtt{\sim}$}}

\begin{document}

\title{Stealthy Hyperuniform Collagen Arrays Impart Structural Color to Bird Skin}

\author{Allison Culbert}
\affiliation{Department of Physics, Yale University}

\author{Halona Dantes}
\affiliation{Department of Physics, Yale University}

\author{Mahathi Narayanaswamy}
\affiliation{Department of Physics, Yale University}

\author{Amanda L. Holt}
\affiliation{Department of Physics, Yale University}

\author{Haina Wang}
\affiliation{Department of Physics, University of Pennsylvania}

\author{Alison M. Sweeney}
\affiliation{Departments of Physics and Ecology \& Evolutionary Biology, Yale University}

\date{\today}

\begin{abstract}
We find that the collagen arrays in brightly colored bird skin exhibit disordered stealthy hyperuniformity of the constituent collagen fibrils. These disordered hyperuniform structures in turn impart bright structural color, in some cases generating clean band gaps. Inverse thermodynamic analysis of micrographs of these systems shows the collagen fibrils exhibit a local soft repulsive potential coupled to an effective decaying oscillatory potential, as predicted by hyperuniformity theory. A larger exploration of parameter space using these effective potentials in molecular dynamics simulations reveals a complex phase space involving liquid re-entrance and the ability to make disordered materials with bright reflectance across the entire human-visible spectrum. Surprisingly, we found that most of the native living systems likely have an intense reflectance peak in the ultraviolet, where birds have sensitive color vision, while humans perceive the long-wavelength shoulder of this peak $>400\ nm$ as blue. Unjammed soft disks in real systems such as colloids likely cannot realize the long-range features of the effective two-dimensional potentials required to achieve these optical effects. However, we observe that the three-dimensional nature of the living systems, which are constituted of rod-like fibrils packed with a soft local repulsion and small chiral pitch, could generate the effective long-range features of the extracted two-dimensional potentials that generate bright reflectance in these disordered system. To our knowledge, this is the first demonstrated example of disordered hyperuniform order in natural materials producing vivid optical phenomena.\end{abstract}

\maketitle
\section{\label{sec:intro}Introduction:}

Hyperuniform systems are those in which density fluctuations are suppressed at large length scales. Crystalline systems are a special case of hyperuniformity, however hyperuniform materials are often disordered structures. \cite{Torquato2003j.PhysRevE.68.041113} These disordered systems include jammed, athermal soft-sphere models of granular media\cite{Berthier2011,Silbert2009}; maximally random jammed hard-particle packings\cite{Donev2005,Zachary2011,Klatt2016,Atkinson2016}; and some non-equilibrium phase transitions\cite{Jack2015,Hexner2015,Weijs2015,Tjhung2015,Weijs2017,Hexner2017,Kwon2017,Torquato2018j.physrep.2018.03.001}.  

Hyperuniform systems have also been shown to exhibit interesting material and optical properties, including producing complete photonic band gaps (PBGs) \cite{Man2013}. Previously, it was believed that only crystalline structures (known as photonic crystals (PhCs)) were capable of producing PBGs \cite{Yablonovitch1987, Florescu2009}. Therefore, the potential to create soft, disordered materials that avoid the drawbacks of crystalline materials such as small size and fragility while also exhibiting material and transport properties such as complete PBGs and lossless waveguiding is of particular interest \cite{man2013isotropic,Man2013}.

The field of disordered optics was initially motivated by understanding and suppressing the significant effects of irregularities in crystalline structures on their optical properties \cite{Yablonovitch1987, PhysRevLett.58.2486}. However, researchers have shown that disordered structures have useful optical properties, and can also be made at a relatively large scale and low cost. Some applications of disordered optics include anti-reflective coatings \cite{Minot:76}, thin-film photovoltaic cells \cite{Deckman1983}, low-emissivity coatings \cite{Glaser:08}, and surface-enhanced Raman spectroscopy sensors \cite{Nishijima:13, Lalanne:25}.  In this context, hyperuniformity theory provides a powerful framework to design disordered photonic materials of specific function, as well as to learn the inter-particle potentials that would allow for self-assembly of these materials in engineered systems \cite{Zhang2015,wang2024designer}. 

There are a few documented examples of hyperuniform organization in nature, such as the organization of photoreceptors in chicken retina \cite{Jiao2014j.PhysRevE.89.022721}, receptors in the immune system \cite{Mayer2015}, and the distribution of vegetation in drylands \cite{Hu2025}. However, these are not materials in the traditional sense, as the physics governing their functions obey system-specific biological laws rather than thermal or chemical potentials. 

\begin{figure}[tbp]
     \centering
     \includegraphics[width=\textwidth]{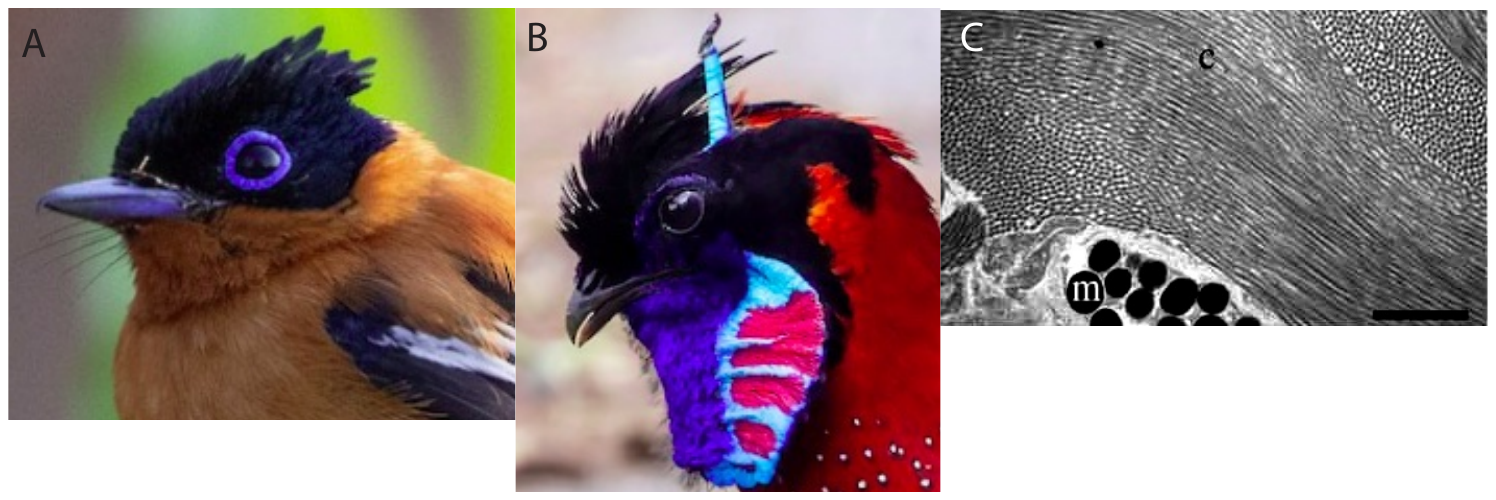}
     \vspace{-190 pt}
         \caption{\textbf{Bird skin with structurally colored collagen fibril arrays} \textbf{A)} \textit{Terpsiphone mutata} with violet-colored periorbital skin which is visible as a bright eye-ring (system viii below). Image is obtained from Macaulay Library (Asset no. ML501994851) with copyright belonging to Ian Davies. \textbf{B)} \textit{Tragopan satyra} with several structural colors in its wattle (system vi below). Image is obtained from Macaulay Library (Asset no. ML566125371) with permission to publish from Bradley R. Hackers. \textbf{C)} TEM of large-area view of collagen fibril arrays from \textit{Tragopan caboti} showing three-dimensional helical pitch of fibrils living system. Scale bar is 2 $\mu m$. Image reproduced with permission from Fig.~6 of Prum and Torres~\cite{Prum2003jeb.00431},
\textit{J.\ Exp.\ Biol.} \textbf{206}(14), 2409--2429 (2003),
\url{https://doi.org/10.1242/jeb.00431}.
Used with permission of The Company of Biologists Ltd, from ``Structural colouration of avian skin: convergent evolution of coherently scattering dermal collagen arrays,'' R.\ O.\ Prum and R.\ Torres, vol.\ 206, 2003; permission conveyed through Copyright Clearance Center, Inc.}
         \label{fig:fig0}

\end{figure}

To investigate whether evolution may leverage the principles of hyperuniformity to create materials with novel optical functions, we turn to structurally colored bird skin. Many species of birds exhibit structural color in exposed regions of skin that arises from the arrangement of the skin's constituent collagen fibrils (Fig. \ref{fig:fig0}A,B \cite{Prum2003jeb.00431}). We observed that the arrays of collagen fibrils composing this brightly colored skin are often highly disordered, with patterns of local packing that are qualitatively liquid-like rather than crystalline systems with defects or systems that have jammed (Fig. \ref{fig:fig1} i, iii-x). The collagen arrays in bird skin generally produce bright, matte, isotropic colors, in contrast to the shinier, directional colors associated with many iridescent feathers. 

Here we present a re-analysis and analytical extension of transmission electron microscopy data (TEM) images reported in \cite{Prum2003jeb.00431}. Our work successfully rationalizes the optical properties of structurally colored avian skin as arising from band gaps produced by disordered hyperuniform order of collagen fibril arrays. We find that in several species, these arrays exhibit different degrees of so-called stealthy hyperuniformity, in which there is a range of spatial frequencies around the origin for which density fluctuations are effectively zero. Many of the samples in the TEM dataset were hyperuniform unto themselves; extended physical analysis the spatial statistics of these images suggest that all of the living systems may be hyperuniform. In a given species, the combined parameters of degree of hyperuniformity, fibril diameter, and inter-fibril spacing generates the inter-specific variation in skin color. In general, as the degree of collagen fibril hyperuniformity in a given species' skin increases, the color generated by that species becomes more spectrally pure, i.e., the reflectance peak becomes brighter and narrower, in some cases approaching a perfect band gap. 

As part of our analysis, we used "inverse thermodynamics" to extrapolate a set of two-dimensional effective fibril-fibril pair-potentials responsible for generating disordered hyperuniform band gaps in these species. We found these potentials to be surprisingly similar between species with disordered hyperuniform structural color, consistent with a local soft repulsion and decaying oscillatory character at distances of $\gtrsim$1.5 particle radii, as predicted by hyperuniformity theory \cite{Torquato2018j.physrep.2018.03.001, wang2024designer}. In two dimensions, a soft system of 100-nm scale colloids or rods such as collagen fibrils likely cannot produce the effective long-range potentials estimated from the bird skin systems. However, larger area electron micrographs of the bird skin \cite{Prum2003jeb.00431} show that in small regions, these systems are effectively 2D disks (Fig. \ref{fig:fig1}A), but in larger volumes consist of rods packed with a helical pitch (Fig. \ref{fig:fig0}C). Our effective two-dimensional inverse thermodynamic potentials are consistent with a local soft repulsion coupled to a longer-range set of forces generated out-of-plane within this larger three-dimensional chiral structure. Our simulations also suggest that re-entrant melting may also play an important role in the living systems' ability to generate bright colors.

Our work documents the first example of a set of evolved materials with optical properties that emerge from stealthy hyperuniform organization of the materials' constituents. We also provide a comprehensive analytical and computational framework for assessing similar structures from small datasets such as single TEM images in other natural systems. We also find that, remarkably, the extracted collagen fibril potentials, when explored throughout a larger parameter space than that directly documented in bird skin, are capable of producing isotropic structural colors of a wide gamut of human-visible hues from red to violet. As such, this work will be important for informing the engineering of environmentally stable, benign pigments and coatings of arbitrary reflectance.

\section{\label{sec:mat-methods}Materials \& Methods: }

\subsection{\label{sec:method1}Image Analysis}

\subsubsection{Binarization and Parameter Calculation}
We analyzed 12 transmission electron microscopy (TEM) images of collagen fibril arrays from structurally colored bird skin previously published by Prum and Torres in 2003\cite{Prum2003jeb.00431}; this study was in turn a comprehensive study of this type of structural color in birds. The species and tissue in each image relative to the lower-case roman numeral index we use for each of these systems in this work are listed in Table \ref{tab:tableS1}. We binarized the images from this publication using Meta's Segment Anything Method (SAM) software \cite{kirillov2023segment}. MATLAB function \textbf{regionprops} was used to find the centers of particles correctly segmented by SAM. Particle centers missed or incorrectly by SAM were corrected manually.

The resolution of the published images was such that after binarization, some fibrils were improperly joined, thus \textbf{regionprops} would identify two or more fibrils as one object. To isolate single fibrils and measure their radii, a marker-controlled watershed transform \cite{watershedVINCENTandSOILLE, Soille2004SegmentatioandWatershedChapter,Meyer1994Watershedpaper} seeded by the fibril center locations was used. Fibrils at the edge of the images whose boundaries went beyond the field of the image were discarded. After an initial watershed transform, the dimensionless Feret aspect ratio ($\theta_{AR}$) was determined for each identified object \cite{WALTON1948Feret}:
\begin{align}
    \theta_{AR}=\frac{\theta_{max}}{\theta_{min}}
\end{align}
$\theta_{max}$ and $\theta_{min}$ are the maximum and minimum Feret diameters of the object, also calculated using \textbf{regionprops} which has a Feret analysis utility that implements a rotating caliper algorithm \cite{toussaint1983CALIPER}. For ideal disks, $\theta_{AR}\approx1$. Objects exceeding a threshold of $\theta_{AR}>1.5$ were flagged as likely clusters of fibrils and subjected to a second, localized watershed to resolve the merged components. In one TEM image, fibrils were so densely packed that fibril centers were located manually.

Once individual fibrils had been located, $\theta_{max}$ and $\theta_{min}$ were calculated for all individual fibrils. Assuming the fibrils have a disc-like cross-section, the average fibril radius ($F$) was approximated as:
\begin{align}
    F \approx \frac{\theta_{max}+\theta_{min}}{4}
\end{align}

Voronoi diagrams \cite{Aurenhammer1991VORONOI} were constructed of the fibril centroids identified in this analysis using the function $\textbf{locality}.\textbf{Voronoi}$ from the Freud Python package \cite{Ramasubramani2020}. The resulting Voronoi diagrams, which also demonstrate the nature of fibril identification, are shown superimposed over the parent TEMs in \ref{fig:fig1}. 

\subsection{\label{sec:method2}Estimating the degree of hyperuniformity of collagen fibrils in structurally colored bird skin}

Hyperuniformity is the suppression of density fluctuations at large distances \cite{Torquato2018j.physrep.2018.03.001}. There are numerous ways to determine whether a system is hyperuniform. Here we use three metrics: fibril number variance, fibril area fraction variance, and fibril structure factor. These parameters are described at length in \cite{Torquato2018j.physrep.2018.03.001} and summarized below for clarity. Hyperuniform systems can be further classified by their degree of hyperuniformity with parameters $\textbf{H}$ and $\alpha$, also described below \cite{Torquato2018j.physrep.2018.03.001}.

\subsubsection{Number Variance} 
The number variance of collagen fibrils, $\sigma_N^2(R)$, where $N$ is the number of fibril center and $R$ is the radius of a window of analysis, was calculated following the procedure described in Ref. \cite{Torquato2003j.PhysRevE.68.041113}. Random points were picked as window centers and the window radii were increased. The number of fibril centers, $N$, enclosed in each window was counted for the set of windows with radius $R$. We then normalized $R$ by the average fibril diameter $F$ of a given system such that $\tilde{R}=\frac{R}{F}$. The variance of $N$ as a function of normalized window size, $\sigma_N^2(\tilde{R})/(\tilde{R})^2$, was calculated. The relation $\log(\sigma_N^2(\tilde{R})/(\tilde{R})^2)$ vs. $\tilde{R}$ was fitted to a line for $\tilde{R} > 1$. If the slope of the fitted line is negative, the structure is likely hyperuniform \cite{Torquato2018j.physrep.2018.03.001}.  Calculated slopes for each TEM are given in the supplement in Table \ref{tab:tableS1}. 

\subsubsection{Area Fraction Variance}
Area fraction variance of fibrils, $\sigma_{\tau}^2(R)$, was calculated following the method outlined in \cite{Dreyfus2015}. $\sigma_{\tau}^2(R)$ is calculated similarly to $\sigma_{N}^2(R)$, however instead of finding the variance of the number of fibril centers in windows of radius $R$, the variance of the area fraction of fibrils in the window ($\tau$) was calculated. This was done by convolving the binarized TEM images with a kernel of a binarized circle and dividing the resulting array by the area of the kernel. This procedure returns the area fraction for all possible placements of the circle on the image for a range of $R$. Then, $\tilde{R}^2\sigma_{\tau}^2(\tilde{R})$ was calculated and $\log(\tilde{R}^2\sigma_\tau^2(\tilde{R}))$ vs. $\tilde{R}$ was fitted linearly for $\tilde{R}>1$. Similarly to number fraction variance, if the slope of $\log(\tilde{R}^2\sigma_\tau^2(\tilde{R}))$ is negative, the structure is hyperuniform \cite{Torquato2018j.physrep.2018.03.001}. These calculated slopes for each TEM are also given in the supplement in Table \ref{tab:tableS1}. 

\subsubsection{\label{sec:methodalpha}Estimation of $\alpha$}

From $\sigma_N^2(\tilde{R})$, we can calculate the hyperuniformity parameter $\alpha$. $\alpha$ is defined in the relation

\begin{equation}
    \sigma_N^2(\tilde{R}) \sim \tilde{R}^{d - \alpha}
    \label{eq:alpha}
\end{equation}
where $d$ is the dimensionality of the system (in our systems, $d=2$). If $\alpha > 0$, the system is hyperuniform \cite{Torquato2018j.physrep.2018.03.001}. 

$\alpha$ was found by fitting $\log(\sigma_\tau^2(\tilde{R}))$ vs. $\log(\tilde{R})$ linearly using the Python function \textbf{polyfit} from NumPy. Eq. \ref{eq:alpha} denotes that the slope of the linear fit to the loglog curve would be equivalent to $d-\alpha$.

\subsubsection{Structure factor} 
Structure factor, $S(\textbf{k})$, is defined as: 
\begin{equation} 
    S(\textbf{k})=1+\rho \tilde{h}(\textbf{k}) \label{eq:S(k)}
\end{equation}
where $\rho$ is the number density, $\tilde{h}(\textbf{k})$ is the Fourier transform of the total correlation function $h(\textbf{r})=g_2(\textbf{r})-1$, and $g_2(\textbf{r})$ is the pair correlation function of the system. 

In this work, we calculate structure factor using

\begin{equation} 
    S(\textbf{k})= \frac{1}{N}\left| \sum_{j=1}^{N}\text{exp}(i\textbf{k}\cdot  \textbf{r}_j) \right|^2 \label{S(k) comp}
\end{equation}
where $\textbf{k}$ is the wavenumber and $\textbf{r}_{j}$ is the position of the $j_{th}$ particle center \cite{Jiao2014j.PhysRevE.89.022721}. We avoided oversampling our systems by choosing \textbf{k} values from an evenly spaced lattice in \textbf{k}-space.  To calculate the one-dimensional structure factor, these lattice \textbf{k}-values were binned and their $S(\textbf{k})$ values were averaged. 
    From the $S(\textbf{k})$ values, the hyperuniformity metric 
\begin{equation}
    \textbf{H} = \frac{S(\textbf{k=0})}{\max \left(S(\textbf{k})\right)} \label{eq:H}
\end{equation} 
was calculated, such that when $\textbf{H} < 10^{-3}$, the system is defined to be "effectively hyperuniform" \cite{Torquato2018j.physrep.2018.03.001}. 

\subsection{\label{sec:method3} Fibril-fibril Pair-potential Estimation}
The fibril-fibril pair potentials of these systems, $v(\textbf{r})$, were estimated using the Ornstein-Zernike equation with the Hypernetted Chain (HNC) closure \cite{Torquato2022j.PhysRevE.106.044122}. This pair-potential approximation is given by:

\begin{equation} 
    \beta v(\mathbf{r}) = h(\mathbf{r}) - ln[g_2(\mathbf{r})]-c(\mathbf{r})
    \end{equation}    
where $\beta = \frac{1}{kT}$, T is the temperature of the system, and $c(\mathbf{r})$ is the direct correlation function, related to $S(\textbf{k})$ by its Fourier transform:\begin{equation}
  \tilde{c}(\textbf{k})=\frac{\tilde{h}(\textbf{k})}{S(\textbf{k})}
\end{equation}

The HNC approximation works best for monodisperse systems, however, the TEM images in \cite{Prum2003jeb.00431} had varying degrees of fibril radius polydispersity. In some systems, polydispersity seemed to be an intrinsic feature of the system (i, iii, x, xi), while others had vary small variation in fibril diameter (iv-ix, xii). For systems with higher, apparently intrinsic polydispersity, the HNC potentials had features that resulted in overlapping particles. To address this unphysical result, we constructed hard disc (hd) potentials using with radius equal to $F$. We then generated a potential coupling the hard disc and HNC potentials (hd+HNC), such that: 

\begin{equation}
v(r) = 
\begin{cases}
    \infty, &   r \leq F \\
    v_{HNC}, &  r > F
\end{cases}
\end{equation}

This potential prevented unphysical overlap of particles in simulations of more polydisperse systems while performing well in recapitulating structures directly observed in the parent TEM images.

\subsection{\label{sec:method4} Molecular Dynamics Simulations and Simulation Analysis}

\subsubsection{Molecular Dynamics Simulations}
Simulations of structurally-colored bird collagen fibrils were generated using HOOMD-blue molecular dynamics software \cite{Anderson2020j.commatsci.2019.109363}. Code was written in Python, and run using the Yale Center for Research Computing (YCRC) clusters. 

Each simulation of collagen fibrils was run with a kinetic temperature of 2.577 kJ/mol (corresponding to 309 K or 98º F, the approximate temperature of living bird skin \cite{prinzinger1991body}). Each particle's mass was 400 kDa (the same mass as a 1 nm slice of collagen fibril \cite{REVELL2021100079}). 

Initially, the simulation number density (or fibrils per area, $\rho$) was chosen to match the original TEMs. Then, to determine the influence of of $\rho$ on the optical and hyperuniformity properties of the system, we also simulated a sweep of $\rho$ values for each system using both the hd+HNC and hd potentials.  We started these density sweeps at $\rho=1\times10^{-5}$ and increased $\rho$ with increments of $1\times10^{-5}$ to a maximum density, $\rho_{max}$. $\rho_{max}$ for each system was chosen such that the packing fraction, $\phi$, of this simulation approached but did not exceed 0.907 ($\phi$ of a triangular lattice) given the system's fibril radius of $F$. The side length of the simulation box was 8 $\mu$m and periodic boundary conditions were used.  The hd+HNC potential simulations were run for 10 $\mu$s. The hd potential simulations were run for 1 $\mu$s. In general, most simulations equilibrated under these conditions, and details of the equilibration behavior are further explored in results.  

\subsubsection{Simulation Analysis}
$S(\textbf{k})$ was calculated for each step of the simulations using Eq. (\ref{S(k) comp}). $\mathbf{H}$ values were calculated using Eq. (\ref{eq:H}). The mean-squared displacement (MSD) of each simulation was calculated using the \textbf{msd.MSD} function from the Freud python package \cite{Ramasubramani2020}. The linear fit of log(MSD) vs log(t) was calculated to determine whether the motion was Brownian, subdiffusive or superdiffusive at a given time interval.

\subsection{\label{sec:method5} Finite-difference time domain optical modeling}
We modeled reflectance from each of these simulated systems using the Finite-difference time domain (FDTD) method \cite{taflove2005computational} as implemented by MEEP, a freely available software package \cite{oskooi2010meep}. We analyzed the results using MATLAB and a colorimetry model of bird vision as described in \cite{endler2005comparing}. 

\subsubsection{FDTD model definition}
Two-dimensional configurations of discs representing 1-nm thin slices of collagen fibrils were generated by the molecular dynamics software, HOOMD Blue, described above. These disc configurations were used to generate center positions for an array infinitely long fibril cylinders with a refractive index of densely packed protein (1.56) embedded in a material with the refractive index of water, 1.33. The radius of the cylinders in each simulation was set to the average value $F$ measured using the image analysis described above. These cylinders were arranged on a square grid eight $\mu m$ on a side and incident light was modeled using a planar gaussian source. The reported reflectance is an average from two separate runs with two different sources, one polarized along the axis parallel to the fibers ($E_z$) and the other with polarization perpendicular to the axis parallel to the fibers ($H_z$). The source was one $\mu m$ in length, spanned 0.2-1.0 $\mu m$ in wavelength and had a resolution of 100 pixels per $\mu m$. Subsequent flux redistribution was measured using four flux planes 12 $\mu m$ in length arranged in a square concentric with the area containing the cylinders. We also included an absorbing perfectly matched layer (dpml) of one $\mu m$ outside these model components. This arrangement allowed for calculation of the radiance reflected, transmitted, and lost from the structure. The simulation was run until the signal at the transmittance flux plane was less than $10^{-3}$ relative to the initial intensity of the source. Reflectance, transmittance and scattered radiation were calculated by normalizing relative to a simulation of the same system without the cylinders. 

To examine the effects of visible wavelength absorption of water and collagen on the reflectance of the modeled systems, we added dispersion using MEEP's Lorentzian susceptibility profile feature. We used eps\_fit\_lorentzian.py (available through MEEP documentation/Materials Dispersion) to fit known or estimated refractive index ($n$) and extinction coefficient ($k$) values from the literature for both materials. The extinction coefficient for collagen was calculated from reports of the absorption coefficient at and around wavelengths between 300-1000 nm (\cite{Sekar2017,Wang2008}). Two Lorentzian oscillators fit $n$ and $k$ well for both water and collagen. The MEEP Lorentzian susceptibility material property requires frequency, $\gamma$ and $\sigma$ for each oscillator representing the material dispersion. The best fit values used for water were: $f_{h20\_0}=0.3859$, $\gamma_{h20\_0}=0.0087$, $\sigma_{h20\_0}=0.0389$, $f_{h20\_1}=11.6093$, $\gamma_{h20\_1}=0$, and $\sigma_{h20\_1}=0.7599$. The best fit values used for collagen were: $f_{c\_0}=12.2584$, $\gamma_{c\_0}=0.0024, \sigma_{c\_0}=1.0072$ and $f_{c\_1}=0.3875, \gamma_{c\_1}=0.0003$, and $\sigma_{c\_1}=0.0446$.

\subsubsection{Reflectance analysis}
Reflectance data from the FDTD simulations were analyzed using MATLAB. Reflectance peak positions, peak widths (calculated at the half-height of the peak) and peak heights were calculated using the built-in function \textbf{findpeaks}. We define a peak quality $A$ to be the ratio of peak height to peak width of the most intense peak of a given reflectance spectrum. Since birds have four visual receptor cone types including one centered at $\sim$ 380 nm \cite{cuthill2000ultraviolet} (rather than three cones with peaks $>$ 400 nm as in humans\cite{roorda1999arrangement}), we used a perceptual model of bird color vision to quantify how the light reflected from each collagen system is located in bird perceptual color space. Following the procedures of Stoddard et al. \cite{stoddard2008evolution} and Endler and Mielke .\cite{EndlerMielke2005}, and analogous to tristimulus values for human color vision, we sum over the product of the modeled reflectance, ambient light spectra (D65) and average bird cone spectral sensitivity curves \cite{stoddard2008evolution} and normalize by the ambient light spectra then log the resulting sum to get the "photon catch" per cone. The photon catch values can be transformed into a tetrahedral space and represented by $R$, $\theta$ and $\phi$ where $R$ is the saturation of the reflectance curve perceived by a bird. $R$ is the distance from the middle of the tetrahedron (where there is equal photon catch per cone). Because $R$ is limited by the edges of the tetrahedron, we normalize $R$ by its maximum value to obtain the chroma ($AC$).

We also calculate human tristimulus values from the modeled reflectance spectra. For this, we sum the product of reflectance, a reference source (D65) and human sensitivity curves separately red, green and blue. Then we normalize each by the sum of the product of the green sensitivity cone and the reference source. We calculate rgb from the tristimulus values using the MATLAB built-in function \textbf{xyz2rgb}. 

\section{\label{sec:results}Results:}
\begin{figure}[tbp]
     \centering
         \includegraphics[width=\textwidth]{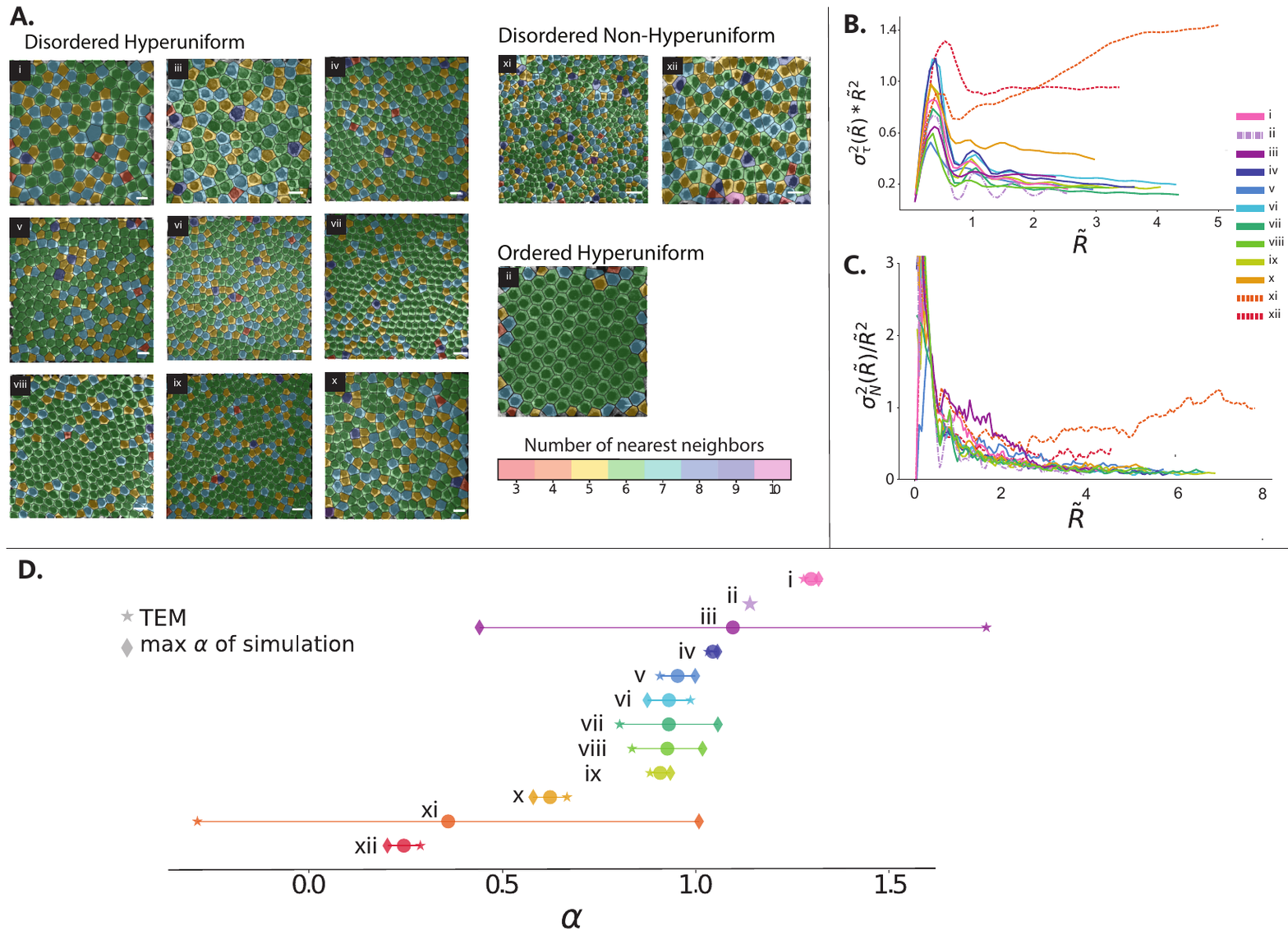}

        \caption{\textbf{Bird skin collagen fibril arrays are generally hyperuniform} \textbf{A)} Voronoi diagrams overlaid on TEMs of bird skin collagen fibrils that are disordered hyperuniform. Species and tissues in each image are listed in Table S1. Area fraction variance of collagen fibrils in \textbf{A}. \textbf{C)} Number fraction variance of collagen fibrils in \textbf{A} \textbf{D)} Parameter $\alpha$ characterizing strength of hyperuniformity of images in \textbf{A} and after molecular dynamics simulation (0$\le\alpha\leq 1$ indicates hyperuniformity, see methods and results below). Throughout this work, collagen fibril TEMs were re-labeled from the original publication, here indexing them to their degree of hyperuniformity as indicated by $\alpha$, with image $i$ having greatest $\alpha$ and the $xii$ smallest, see Sec. \ref{sec:methodalpha} and Table \ref{tab:tableS1} for details. TEM images  in \textbf{A}-\textbf{C} reproduced with permission from Fig.~7 of Prum and Torres~\cite{Prum2003jeb.00431},
\textit{J.\ Exp.\ Biol.} \textbf{206}(14), 2409--2429 (2003),
\url{https://doi.org/10.1242/jeb.00431}.
Used with permission of The Company of Biologists Ltd, from ``Structural colouration of avian skin: convergent evolution of coherently scattering dermal collagen arrays,'' R.\ O.\ Prum and R.\ Torres, vol.\ 206, 2003; permission conveyed through Copyright Clearance Center, Inc.
        \label{fig:fig1}}
\end{figure}

\subsection{\label{sec:result1}Image Analysis}

The Voronoi diagrams of the binarized TEM configurations are displayed over the original TEM images in Fig. \ref{fig:fig1}. The Voronoi diagrams provide a high-fidelity recapitulation of fibril locations and nearest-neighbor relations, verifying that the binarizaiton procedure produced reliable results. Binarized TEM images with located fibril centroids are shown in Fig. \ref{fig:sup1}. 

\subsection{\label{sec:result2}Calculations of the degree of hyperuniformity in literature electron micrographs}

The linear fits of $\log(\sigma_N^2(\tilde{R})/\tilde{R}^2)$ and $\log(\tilde{R}^2\sigma_\tau^2(\tilde{R}))$ from Fig. \ref{fig:fig1} are listed in Table \ref{tab:tableS1}. These results show that collagen fibrils in images i, and iii-x from the species \textit{Numida meleagris, Tragopan satyra, Tragopan caboti, Syrigma sibilatrix, Ramphastos toco}, and \textit{Terpsiphone mutata} are disordered hyperuniform systems. Image ii from \textit{Philepitta castanea} was an ordered hyperuniform system, i.e., crystalline. 

The direct TEM images xi and xii from \textit{Oxyura jamaicensis} and \textit{Gymnopithys leaucapsis} were disordered and non-hyperuniform. The slope of number variance for image xii is near-constant and at a high value such that we would not anticipate it to decay to zero at greater lengths, while both metrics for image xi are plainly increasing. Image iii from \textit{Procinas nudicollis} produced negative slopes in both methods, however the slope of $\log(\tilde{R}^2\sigma_\tau^2(\tilde{R}))$ was -0.03, very close to zero, while $\log(\sigma_N^2(\tilde{R})/\tilde{R}^2)$ was -1.76, steeply negative. However, the linear fits for this system were poor due to fluctuations. The hyperuniformity classification for this system is therefore somewhat ambiguous, but given the negative slope in the simpler number density metric, we group it with the disordered hyperuniform systems.

Structure factors for each TEM were calculated, and are shown in Fig. \ref{fig:sup2}. Most of the structure factors decay towards 0 for $k \rightarrow0$. Three outliers are the same images mentioned above: iii, xi, and xii. These exhibit upturns in S(k) as $k \rightarrow 0$. Given the inverse space nature of the structure factor, and the loss of small k data for small TEM dimensions, S(k) analysis was less reliable for predicting hyperuniformity in these images. However their agreement with real space analysis was encouraging.

\begin{figure}[tbp]
     \centering
     
         \includegraphics[width=\textwidth]{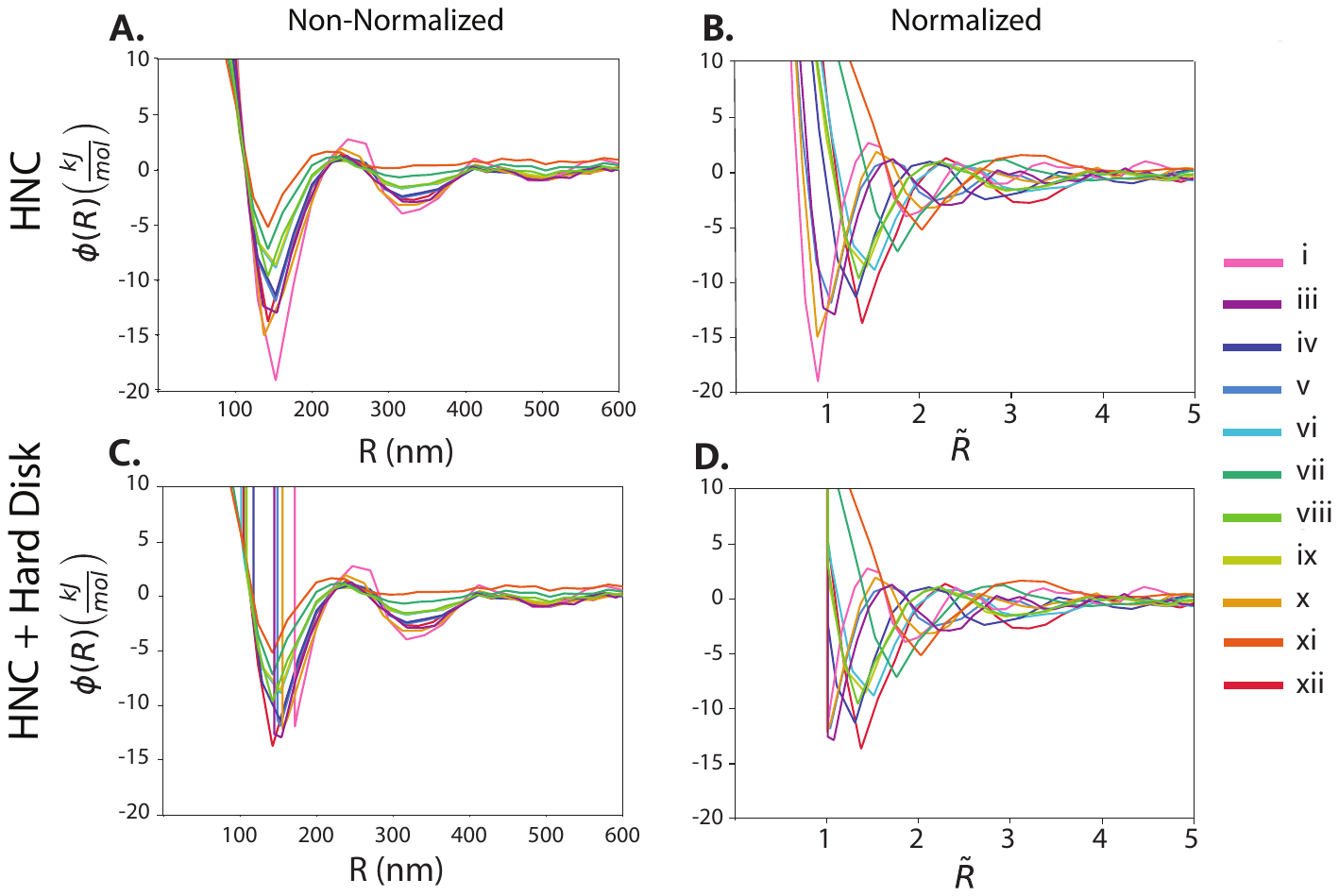}
         \caption{\textbf{Effective fibril-fibril pair potentials.} \textbf{A)} HNC potentials for the 11 disordered collagen fibril systems in Fig. \ref{fig:fig1} as estimated using methods in \ref{sec:method3}. \textbf{B)} Same potentials as in \textbf{A},  normalized by characteristic fibril radius $F$ for each system. \textbf{C)} The HNC potentials shown in \textbf{A} coupled to a hard-sphere potential consistent with the average fibril radius $F$ calculated for each system. \textbf{D} Same potentials as in \textbf{C}, normalized by characteristic fibril radius $F$ for each system.}
         \label{fig:HNC_potentials}

\end{figure}

\subsection{\label{sec:result3} HNC potential generation}

Hyperuniformity is a measure of order at large distances. However, the TEMs from \cite{Prum2003jeb.00431} analyzed here are a mere 4 µm$^2$ in area. We found that ten of the twelve collagen arrays from these images have suppressed density fluctuations at the largest length scales of the image. However, two technical issues remained: the longest lengths and corresponding number of particles represented in the images were too small for definitive quantification of hyperuniformity, and the literal structures were too small for robust optical modeling in the visible regime. However, these technical issues also presented an opportunity to more carefully probe the thermodynamic nature of these systems and the physical origins of their disordered structural color. We therefore queried whether and how the effective pair potentials of the observed living systems are consistent with hyperuniformity theory; whether the disordered hyperuniform structure observed in nine of the systems is responsible for the bright structural colors of the living skin; and how these effective potentials might in turn generate a larger class of hyperuniform optical materials. 

We estimated effective pair-potentials of collagen fibrils in each bird skin system and used these potentials to generate simulations of bird skin that were larger than the original TEMs for further optical and hyperuniformity analysis. In order to simulate these materials, we required pair potentials that matched the statistics of the original TEM images.

We employed a strategy of "inverse thermodynamics" \cite{Torquato2022j.PhysRevE.106.044122, wang2024designer} to generate systems with enough particles and area for robust structural and optical analysis. Following the procedure outlined in \cite{wang2024designer}, we approximated the pair potentials of the particles in the original images for the purpose of simulating a larger, computationally modeled array of bird skin collagen fibrils thermodynamically consistent with the experimental images.

The HNC approximation resulted in the potentials shown in Fig. \ref{fig:HNC_potentials}A. Fig \ref{fig:HNC_potentials}B is the same potential, with the $x$-axis normalized by the average radius of the particles in a given image. Given the crystalline structure of system ii, the HNC approximation is unable to produce a useful potential. This is due to the Bragg peaks present in $g(\textbf{r})$, which result in $\ln(0)$ terms in the HNC approximation. Therefore, we include eleven HNC potentials in our analyses going forward.

All eleven of the non-normalized HNC potentials are remarkably similar to one another (Fig. \ref{fig:HNC_potentials}A). They exhibit a soft repulsive region from 0 to $\sim$150 nm, and an attractive region spanning from $\sim$150 nm to $\sim$250 nm followed by 1–2 periods of decaying oscillatory potential out to 600 nm. Unexpectedly, normalizing the potentials by the average diameter of particles in a given image markedly stretches or compresses the potentials relative to one another along the axis of radius (Fig. \ref{fig:HNC_potentials}C,D). In these normalized potentials, the attractive region begins somewhere between one to two particle diameters from the particle center, while in the non-normalized potential, the attractive region always begins around 150 nm. This suggests that the interparticle distances in each living system are not determined by the observed particle radii alone, but instead by a particle radius plus a second distance specific to each tissue type, such that the \textit{sum} of these two distances is constant across the 11 samples in this dataset.

The HNC approximation is derived solely from the particle center coordinates in the experimental images from (Fig. \ref{fig:sup1}, \cite{Prum2003jeb.00431}, and does not explicitly incorporate particle radius \cite{Torquato2022j.PhysRevE.106.044122}. Therefore, the repulsion strength at distance $R$ corresponds to the likelihood that particles are a distance $R$ apart. If the original image has even a small number of particles with radii smaller than the average radius, and those particles are a distance $<F$ apart, the potential will reflect this by being softly repulsive instead of infinitely repulsive at that inter-particle distance. 

In our analysis, the TEM images with small fibril radius standard deviation resulted in HNC simulations qualitatively similar to the original systems. For systems with higher degrees of fibril radius polydispersity, simulations with the direct HNC potentials contained artifacts such as exactly overlapping particles and phase separation not seen in the original images. In these polydisperse systems, the HNC potential finds unrealistic forces at small $R$ values, such as being too weakly repulsive or having shallow local minima such that particles get "stuck" at interparticle distances $<F$. 

To address these artifacts, we coupled the HNC potential to a hard-sphere potential for the average radius of a particle in a given system (hd+HNC), Fig. \ref{fig:HNC_potentials}C); all simulations reported here use the hd+HNC potentials in Fig. \ref{fig:HNC_potentials}C.

\begin{figure}[tbp]
     \centering
         \includegraphics[width=\textwidth]{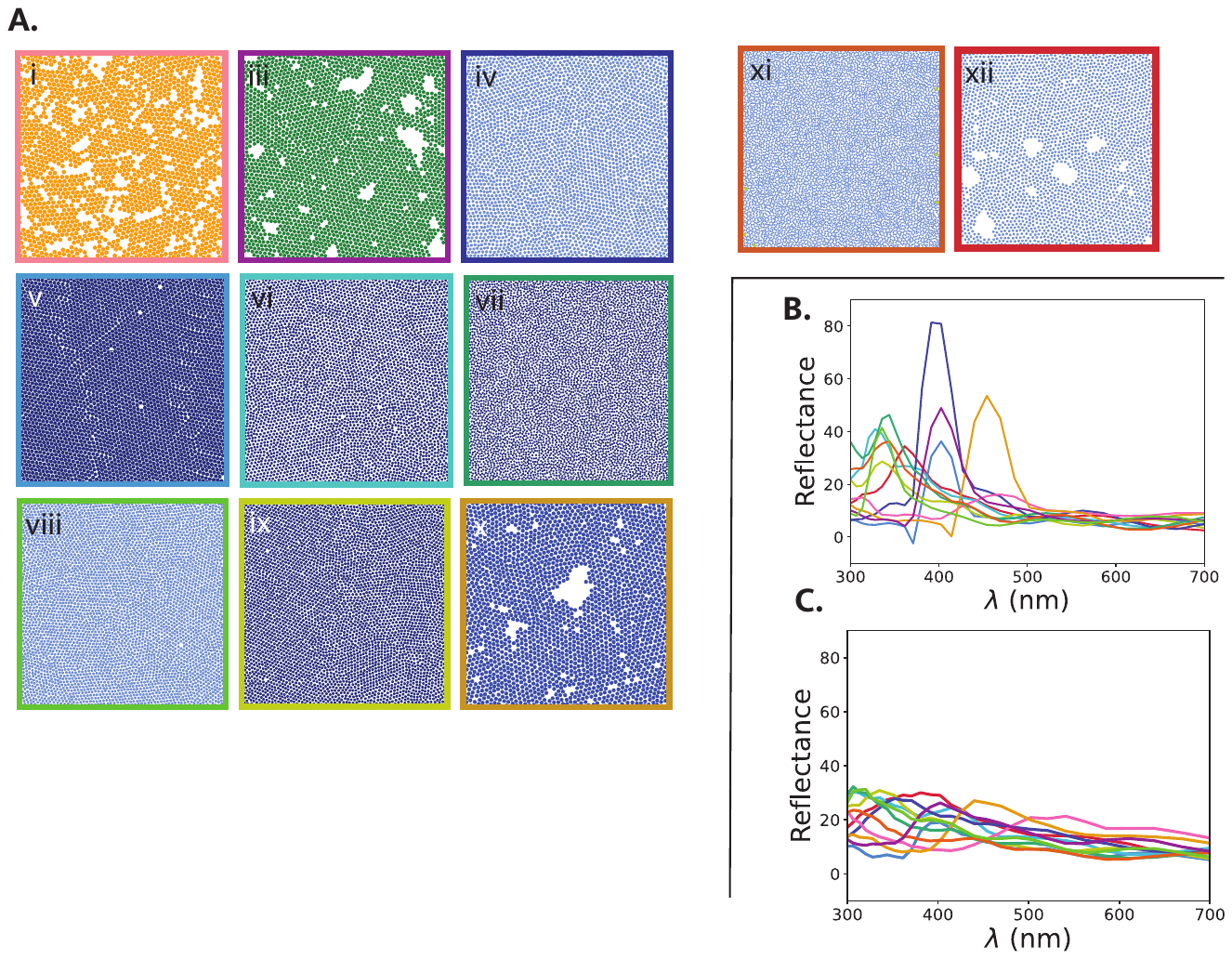}
         \vspace{-60pt}
        \caption{\textbf{MD simulation results for hd+HNC potentials with native values of $\rho$.} \textbf{A)} Final fibril disc configurations from MD simulations with hd+HNC potntials. Box sides are 8 $\mu m$ long and box border colors represent the system symbols carried from Fig. \ref{fig:fig1}B–D. The fibrils are colored by to the human-perceived skin color of the corresponding living bird species. See Fig. \ref{fig:sup5} for corresponding configurations using hd potential. \textbf{B)} Calculated reflectance of the fibril structures in \textbf{A}, reflectance curves correspond to box border colors in \textbf{A}. \textbf{C)} Reflectance curves for MD simulations of matched $F$ and $\rho$ for systems in \textbf{A} but using the hd potential. Reflectances are markedly lower and broader when particles have the same size and density but have short-range potentials. }\label{fig:fig3}
\end{figure}

\subsection{\label{sec:result4} Simulations and Analysis}

Three main groups of molecular dynamics (MD) simulations were run: original parameters from TEM images, a sweep in $\rho$, and varying system size. 

\subsubsection{MD using TEM Parameters}
Simulations were initially run with the same $\rho$ values as the TEM images and particles interacted with the hd+HNC potentials. The final frames of these simulations are shown in \ref{fig:fig3}. 

Under these conditions, two of the systems were clearly disordered hyperuniform with $\textbf{H}< 10^{-3}$ (iv, viii); one was at sufficiently high $\rho$ to crystallize, making it ordered hyperuniform (v); five had $\mathbf{H}$ values on the order of $10^{-3}$ making an assignment of hyperuniformity ambiguous (vi, vii, ix, x, xi); and three had $\mathbf{H}$ on the order of $10^{-2}$, making them simply disordered (iii, xii). 

The parameter $\textbf{H}$ depends solely on the absolute value of an extrapolated estimate of $S(\mathbf{k} = 0)$ relative to $S_{max}(\mathbf{k})$, such that in experimental or numerical systems such as these, the accuracy of $\textbf{H}$ is constrained by the dimensions of the simulation. Another way to quantify hyperuniformity is using the parameter $\alpha$, which characterizes the trend of $\sigma^2_N(\tilde{R})$ as $\tilde{R}$ increases, making it less sensitive to the absolute size of the system. As described in section \ref{sec:method2}, the system is hyperuniform if $\alpha>0$. As described in Ref. \cite{Torquato2018j.physrep.2018.03.001}, if the system can be characterized close to the limit $R \rightarrow \infty$, $\alpha$ cannot be $> 1$. In these systems of finite size, we are far enough from the $R\rightarrow\infty$ limit  that $\sigma^2_N(\tilde{R})$ is still oscillating near 1 at the largest experimentally reached values of $R$. Therefore we occasionally find the numerical result that $\alpha\gtrsim1$, that with infinitely large systems may plausibly smooth to $\alpha=1$. 

\subsubsection{MD sweep of fibril density ($\rho$) variation}

\begin{figure}[tbp]
         \centering
         \includegraphics[width=\textwidth]{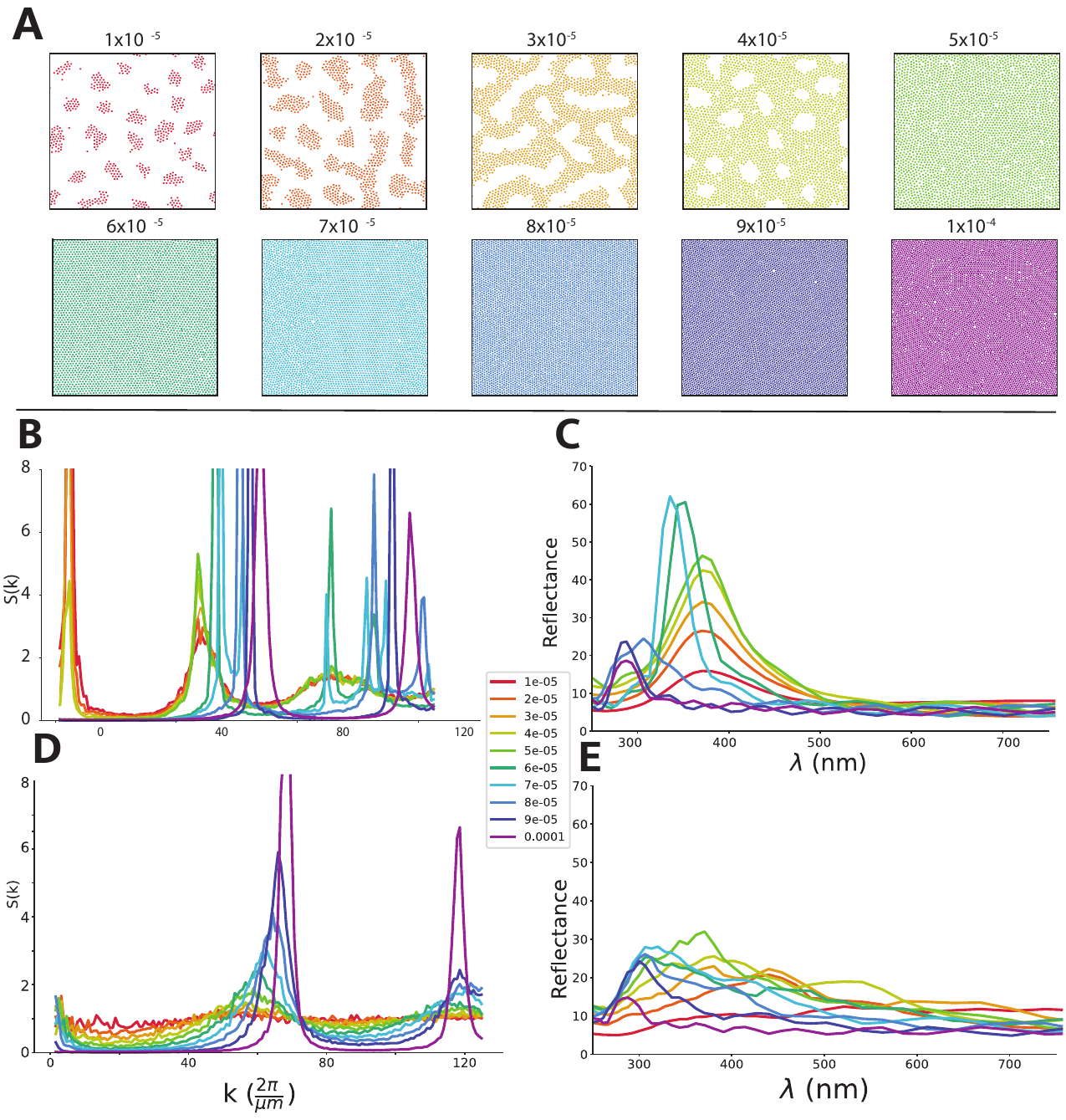}
         \vspace{-120pt}
         \caption{\textbf{Density ($\rho$) sweep of hd+HNC potential estimated from system viii.} \textbf{A)} Final simulation frames as a function of system density. The sparsest densities start as a spinodally decomposed liquid-gas phase, undergo a phase transition to liquid near $5\times 10^{-5}$ fibrils nm$^{-1}$, crystallize near $6\times 10^{-5}$ fibrils nm$^{-1}$ and then exhibit re-entrant melting near $7\times 10^{-5}$ fibrils nm$^{-1}$. This re-entrant density is also the location of the brightest, narrowest reflectance (see panel \textbf{C}). \textbf{B)} Structure factors for fibril configurations in \textbf{A} showing a liquid state and re-entrant melting. \textbf{C)} Reflectance spectra associated with \textbf{A} and \textbf{B}. \textbf{D)} Structure factors of matched systems using the hd potential. \textbf{E)} Reflectance spectra of matched system using the hd potential.}
        \label{fig:fig4}
\end{figure}

The TEMs we used to estimate pair-potentials were small samples of much larger systems. Thus, the $\rho$ found in the small area of a given TEM may not have been the average $\rho$ for the larger system. We are also interested in the behavior of the effective potentials of the living systems in parameter space outside that used by evolution. Therefore, we investigated whether varying $\rho$ while maintaining the same inter-particle potential influenced the system's phase and $\mathbf{H}$ value. To do this, we ran molecular dynamics simulations for the 11 systems with hd+HNC potentials for $\rho$ values ranging from $1\times 10^{-5}$–$1.4\times 10^{-4}$ fibrils $nm^{-2}$. The final frame for each $\rho$ sweep simulation using potential viii is shown in Fig. \ref{fig:fig4}A, and the rest of the systems' $\rho$ sweeps are shown in Fig. \ref{fig:sup4}. 
 
In this exploration of $\rho$, seven of the eleven disordered systems reached a state of disordered hyperuniformity where $\textbf{H}<10^{-3}$ without jamming. The remaining four systems, i, v, x, and xi reached states other than disordered hyperuniform in a density sweep, and each system had distinct features (Fig. \ref{fig:fig4}B). 

In general, as our modeled systems evolved in time, the parameter $\mathbf{H}$ characterizing the degree of stealthy hyperuniformity increased along with the parameter $A$, or aspect ratio of the primary reflectance peak (Fig. \ref{fig:fig5}). This increase was rapid at early time points and plateaued as the system equilibrated. This correlation suggests that the spectrally pure reflectances of these disordered systems results from their disordered hyperuniform character. 

The TEM image of v shows a system that is jammed, consistent with a system of hard discs at high $\rho$ without a long-range pair-potential. 

The system in xi was one for which the parent TEM was disordered but not hyperuniform. In this case, simulations using the potential derived from this image were similarly non-hyperuniform at any density. In contrast, system xii, the other non-hyperuniform parent TEM, has a relatively low native $\rho$ of $4.5\times 10^{-5}$. Our simulations using the hd+HNC potential derived from xii reached hyperuniformity for all $\rho \geq5 \times10^{-5}$. This result suggests that xii is also likely disordered hyperuniform in the living system and that the small area of the published TEM may capture an uncharacteristically sparse region of the tissue. 

Estimating pair potentials via the HNC method cannot explicitly account for polydispersity in particle radius. Systems i and x have the most polydisperse fibril radius $F$ in the set of 11 systems, and correspondingly, these simulations were less successful in producing structures very similar to those of the original images.

We also observed re-entrant phase transitions for five of the systems (vi, vii, viii, xi, and xii); four of these have at least one density that is hyperuniform (vi, vii, viii, and xii), and one is disordered at all densities (xi). When $\rho<5\times10^{-5}$, we observed spinodal-type structures. Then, when $5\times10^{-5} < \rho < 6\times10^{-5}$, these systems underwent a phase transition to volume-spanning liquids and hyperuniformity emerged in vi, vii, viii, and xii. As $\rho$ increased beyond $6\times 10^{-5}$, we observed a crystalline state that then re-entered a liquid and disordered hyperuniform state near a density of $7\times10^{-5}$. The system remained in this state until reaching the maximum possible $\rho$ value for the system and re-crystallizing (example shown in for system viii Fig. \ref{fig:fig4}, other re-entrant systems shown in Fig. \ref{fig:sup4}). The structure factors of many of these systems also show flat $S(\textbf{k})$ with $S(\textbf{k})\sim 0$ for a range of $\textbf{k}$ extending from the origin, consistent with the definition of stealthy hyperuniformity (example in Fig. \ref{fig:fig4}B).

In contrast, a set of control simulations using particles of the same diameters as the native system but using the hd potential resulted in very different structures.  The final frames from these simulations are shown in Fig. \ref{fig:sup5}. These simulations were mostly purely disordered ($\textbf{H}>10^{-3}$). Three simulations reached the threshold $\textbf{H} < 10^{-3}$: vi at $ \rho = 11\times10^{-5}$, ix at $9\times10^{-5}$, and iii at $5\times10^{-5}$. However, jamming can occur at $\phi \geq 0.802$, and the simulations that reached hyperuniformity via the hd potential occurred at $\phi = 0.81, 0.89$ and $0.82$, respectively. We interpret these results to mean that any hyperuniformity in these hd-potential systems emerged from hard disk jamming or crystallization rather than from the long-range portion of the potential as is the case for the HNC+hd simulations described above.

The MSD for each simulation was calculated, and the slope, m, of the log(MSD) vs log(t) plots was calculated over given time intervals, and are shown in Fig. \ref{fig:sup3}. For all systems, $m>1$ for early time intervals. As the system relaxes, the vast majority exhibited Brownian motion ($m = 1$), or subdiffusive states ($m<1$). Some systems exhibited brief periods of superdiffusive motion ($m>1$) and a few rare cases exhibited short periods of ballistic motion ($m\approx 2$). In general, these results show that our simulations allowed for sufficient relaxation or equilibration to support these analysis of their final or relaxed structural states.

\subsection{\label{sec:result5} Finite-difference time domain optical modeling}

FDTD simulations of the MD trajectories were run to find the reflectance spectra of these systems. Each of our simulations using the native TEM parameters produced a reflectance peak in the range of bird-visible wavelengths (300 nm–700 nm) ( Fig. \ref{fig:fig4},\cite{osorio2019evolutionary}).

The original work describing these structural colors reports human-visible colors that are mostly shades of blue with one orange (i) and one green (iii) example. Surprisingly, we found that under the native system parameters, all the other systems produced a more intense peak in the ultraviolet somewhere between 300 and 400 nm (Fig. \ref{fig:fig3}). These results suggest that the blue appearance of these species in human vision is the result of a long-wavelength shoulder of a much brighter peak in the UV, where birds also have color-vision sensitivity, and these disordered hyperuniform optics have evolved as intraspecific signals in the ultraviolet range \cite{endler2005comparing,REVELL2021100079,osorio2019evolutionary}.

Considering the full parameter space of our larger set of simulations including a density sweep, we found a surprisingly large range of hues spanning a large gamut of human and bird perceptual space. These simulated reflectance spectra often have a tall, narrow peak in the ultraviolet region of the spectrum in the region visible to birds, and secondary peaks $>$400 nm in the human and bird visible region.

\begin{figure}[tbp]
     \centering
         \includegraphics[width=\textwidth]{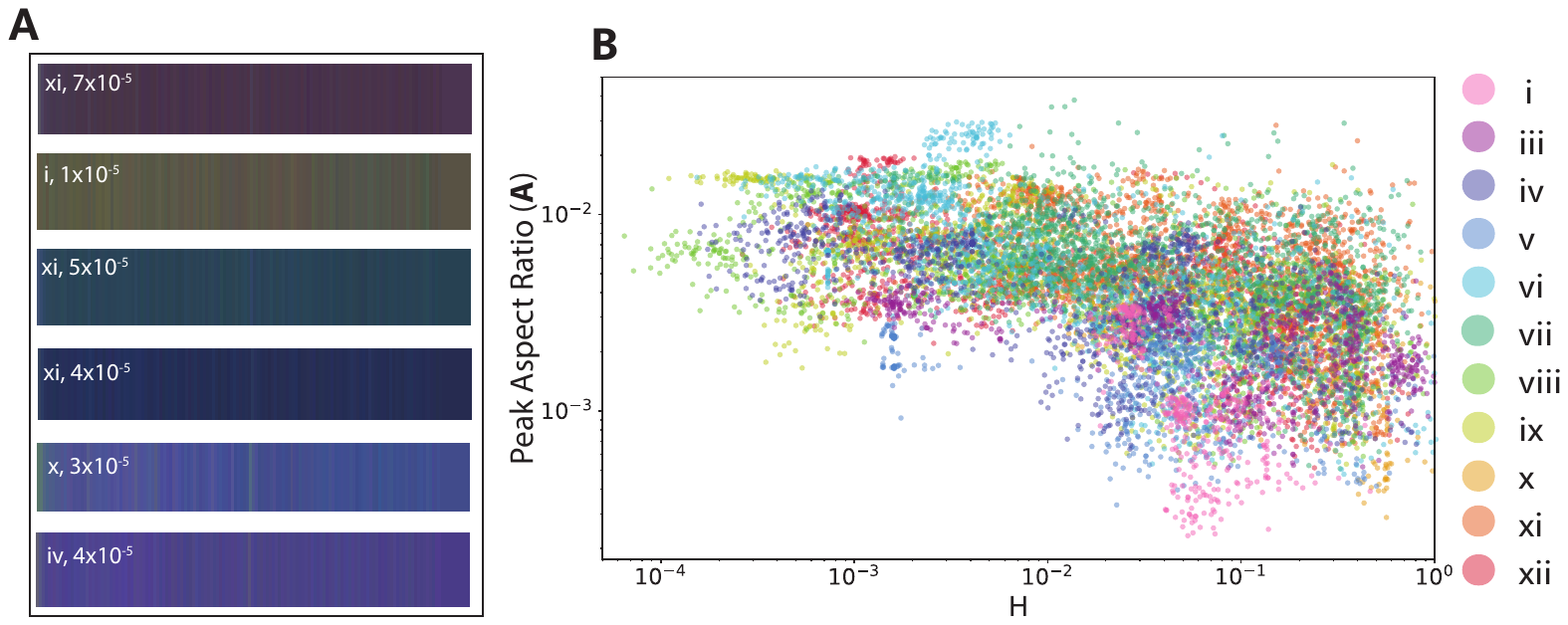}
         \vspace{-100pt}
        \caption{\textbf{Bird collagen fibril systems can achieve arbitrary human-visible hues and achieve narrow, bright reflectance peaks via effective hyperuniformity.} \textbf{A)} Under a variety of parameter combinations of our model, the potentials extrapolated from the bird collagen fibril systems can reflect colors from the human red through violet visible range. Parameter combinations as indicated. Vertical bars in each rectangle represent reflectance from a single frame of the corresponding MD simulation. Simulation proceeds from left to right along a given rectangle. \textbf{B)}. Relation between hyperuniformity parameter $\mathbf{H}$ and reflectance peak aspect ratio $A$ associated with the depth of a bandgap of a given system. Here, colors correspond to species system consistent with the rest of the figures. Several systems reach values of $\mathbf{H}\le 10^{-3}$ and in general there is a negative correlation between $A$ and $\mathbf{H}$.}
        \label{fig:fig5}
\end{figure}

The peak quality $A$ is shown to be negatively correlated with $\mathbf{H}$ for all systems across all $\rho$ values, as depicted in Fig \ref{fig:fig5}B. Thus, as the hyperuniformity of a system increases, its reflectance peak becomes taller and narrower, resulting in a more pure spectra. Additionally, several of our systems reach the threshold of $\mathbf{H}$ of order $10^{-4}$, reaching the strict analytical threshold for defining effective hyperuniformity \cite{Torquato2018j.physrep.2018.03.001}. In contrast, systems with the same density and size particles interacting with the matching hd potential yields low $A$ values and no cases of $\mathbf{H}$ of order $10^{-4}$. Interestingly, high values of $A$ are possible in the hd-potential systems, and occur when $\rho$ is high enough that the system jams or crystallizes, though $\mathbf{H}$ is not effectively hyperuniform.

\section{\label{sec:discussion}Discussion: }

\subsection{Phase Transitions and Reentrant Melting}
In our initial run of simulations at the original $\rho$ values found in the TEMs, we saw a wide range of phases. At smaller $\rho$ values, the particles formed spinodal structures, and also exhibited gas-liquid phase separation. For higher $\rho$ values, we saw volume spanning liquids as well as square and triangular lattice solids. The wide range of phases signaled to us that these pair-potentials may contain interesting phases for varied $\rho$ values.

The re-entrant melting we observed was interesting and unexpected. For these structures, the highest average reflectance peak occurs at $\rho$ either $6\times10^{-5}$ or $ 7\times10^{-5}$, when the structures were no longer crystalline and undergoing re-entrance. This implies that this phase transition from crystalline to a mixture of liquid and crystal phases plays a significant role in the optics of these systems.  The crystals that occur at slightly higher and slightly lower densities are both effectively hyperuniform as they are ordered. The reentrant liquid keeps the long-range oscillations low (so retains the hyperuniformity) but randomizes the short-range coordination structure so that the configuration can be disordered. Therefore we predict that in warm living systems, constructing soft repulsive potentials that create bright color at re-entrant melting densities is a viable way to achieve thermally stable and effectively disordered hyperuniform structures.

\subsection{Collagen fibril structure and the emergence of hyperuniform effective pair-potentials}
The similarity of the effective inter-particle potentials between systems and their divergence upon normalization (Fig. \ref{fig:HNC_potentials}) merits further consideration. Vertebrate collagen fibrils are dense packings of order $10^3$ individual collagen proteins bound in the $\sim$100 nm diameter fibrils observed in TEM \cite{orgel2006microfibrillar}. The external surfaces of fibrils are further decorated by a covalently bound corona of negatively charged polysaccharides \cite{wang2018fucosylated}. We predict that the effective potentials we characterize here arise from a hard, high-index core of densely packed collagen protein, a soft local repulsion due to this low-index polysaccharide corona, and then long-range forces generated out-of-plane from the low chiral pitch of fibril packing in three dimensions. Fascinatingly, while the radius of the high-index fibrils varies between species, the normalized HNC potentials translate apart from each other, suggesting that the size of the repulsive polysaccharide corona varies inversely with the radius of the protein fibril. In other words, the sum of these two distances, the fibril radius and the repulsive distance created by the polysaccharide corona appear to be near-constant across species. These systems may all be governed by the same inter-particle potential, and evolutionary processes apparently tune the system fibril density and fibril radius to change perceived color. The nature of packing and the resulting chiral pitch may also play a role in the final effective potential. We observe that the same system can have reflectance peaks 100 nm apart for different $\rho$ values while maintaining the same fibril radius (Fig. \ref{fig:fig4}C).  This is consistent with several of the birds in which a pale blue and a saturated blue of the same hue are present; pale blue regions are potentially less dense regions of the same or similar supramolecular collagen fibrils.

\subsection{Hyperuniformity vs hard disc}
In the work that originally reported the TEMs analyzed here, Prum and Torres describe color from these arrays as rising from coherent scattering which occurs "when the spatial distribution of scatterers is non-random with respect to the wavelengths of visible light, so that the phases of scattered waves are non-random"\cite{Prum2003jeb.00431}, a concept first described using this term by Bohren and Huffman for interactions of light among larger particles rather than interactions of light with atoms \cite{bohren2008absorption}.

Our results that coherent scattering is a partial description of the optical phenomena observed in bird skin and that the full physical framework of hyperuniformity is required to fully rationalize the remarkable optical effects achieved in these systems.

The difference between coherent scattering and hyperuniformity is most apparent in comparison of optical simulations of the effective bird potentials (hd+HNC) vs. their hard disc (hs) counterparts. If coherent scattering were sufficient to explain these systems, and suppression of density fluctuations at long lengthscales were unimportant, we would expect our hd and hd+HNC results to match. However, we find that our hd simulations result in low-intensity, unststructured reflectance spectra, whereas their hd+HNC counterparts have bright, well-defined peaks that result in highly saturated structural color (Figs. \ref{fig:fig3}B,C; \ref{fig:fig4}C,E). 

In the systems of a given hd+HNC potential and varying $\rho$, hyperuniformity "turns on" around $\rho = 5\times10^{-5}$, which coincides with the onset of narrow, bright reflectance peaks. In total, these observations suggest that evolution appears to select for densities of reentrant melting so that $S(\textbf{k})$ at small $\textbf{k}$ is low, such that these systems have the large-scale-fluctuations of two crystals that both have low $S(\textbf{k})$ but also provides for disorder at short ranges, since the system at those densities must compete between two coordination structures.

We also observe that many systems with $10^{-3} < \textbf{H} < 10^{-2}$ are also capable of producing  saturated structural color (Fig. \ref{fig:fig5}B), leading us to ask what degree of hyperuniformity is required to produce interesting optics. The threshold of $\textbf{H}<10^{-3}$ was set for discussions on "infinitely" large, computationally constructed systems \cite{Torquato2018j.physrep.2018.03.001}. Little work has been done on hyperuniformity in living systems, and research that has been done often focuses on number variance as the metric by which hyperuniformity is achieved, or discuss the trend of S(k) for small k without calculating $\mathbf{H}$ \cite{Liu2024, Mayer2015, Hu2025}. There is currently one documented exploration of three-dimensionally hyperuniform optical systems, however the systems explored were not naturally occurring, and produced band gaps in the microwave regime\cite{Siedentop2024}. We speculate that the three-dimensional character of the bird skin systems produces additional suppression of density fluctuation not captured in our quasi-2D models here, and further exploration into the 3D nature of these materials will provide greater insight into understanding the potential of 3D hyperuniform photonic band gap materials 

In summary, we have documented the first naturally occurring hyperuniform optical materials, calculated the effective two-dimensional potentials underpinning their thermal assembly, shown that these systems across a range of densities can produce an arbitrary array of human- and bird-visible hues, and are often capable of achieving the precise analytical criterion for effective hyperuniformity. We also put forward a model of how these effective two-dimensional potentials could emerge from the three-dimensional system of polysaccharide-decorated collagen fibrils present in the living systems from which we derive this analysis.

\begin{acknowledgments}
We gratefully acknowledge support from the Gordon \& Betty Moore Foundation Experimental Physics Initiative to A.M.S., Yale University, and an NSF Graduate Research Fellowship to H.D. We also thank the Yale Center for Research Computing, specifically Dr. Aya Nawano and Dr. Thomas Langford, for guidance with computation on Yale's Grace High Performance Computing cluster.
\end{acknowledgments}

\bibliography{refs}

\newcommand{\beginsupplement}{%
  \setcounter{table}{0}
  \renewcommand{\thetable}{S\arabic{table}}%
  \setcounter{figure}{0}
  \renewcommand{\thefigure}{S\arabic{figure}}%
  \setcounter{equation}{0}
  \renewcommand{\theequation}{S\arabic{equation}}%
}

\clearpage
\beginsupplement
\section*{Supplementary Information}

\begin{table}[H]
\centering
\begin{tabular}{|c|c|c|c|}
\hline
\rule[-6pt]{0pt}{18pt}System & Species and tissue & Slope of $\log(\sigma_N^2(\tilde{R}) / \tilde{R}^2)$& Slope of $\log(\sigma_{\tau}^2(\tilde{R}) * \tilde{R}^2)$\\ \hline
\romannumeral 1 &\textit{Tragopan caboti} orange& -1.25& -0.71\\
\romannumeral 2 &\textit{Terpsiphone mutata} dark blue & -1.10& -0.82\\
\romannumeral 3 &\textit{Procnias nudicollis} green& -1.76 & -0.98\\
\romannumeral 4 &\textit{Tragopon caboti} light blue & -1.01& -0.65 \\
\romannumeral 5        &\textit{Numida meleagris} dark blue&    -0.91      &    -0.57\\
\romannumeral 6 &\textit{Tragopon satyra} dark blue& -0.97& -0.64\\
\romannumeral 7 &\textit{Ramphastos toco} dark blue& -0.81 & -0.43 \\
\romannumeral 8 &\textit{Terpsiphone mutata} dark blue& -0.84 & -0.64\\ 
\romannumeral 9 &\textit{Tragopon caboti} dark blue& -0.90& -0.41\\
\romannumeral 10 &\textit{Syrigma sibilatrix} light blue& -0.67 & -0.95\\
\romannumeral 11 &\textit{Gymnopithys leucapsis} light blue& 0.29 & 0.14\\
\romannumeral 12       &\textit{Oxyura jamaicensis} light blue&    -0.27&    0.04     \\ \hline
\end{tabular}
\caption{Table of slopes for lines fitted to number and area fraction variance as a function of window radius. If a given slope is $< 0$, the structure is likely hyperuniform; if a slope is $ \ge0$, the structure is not hyperuniform. All of the TEM images from \cite{Prum2003jeb.00431} are hyperuniform, except for (\romannumeral 11), which has a positive slope for both number and area fraction variance, and (\romannumeral 12), which has a negative slope for number variance and a small positive slope for area fraction variance, producing an ambiguous result.}\label{tab:tableS1}
\end{table}

\begin{table}[H]
\centering
\begin{tabular}{|c|c|}
\hline
Number density & H value\\ \hline
\num{1e-5}      &    0.1031         \\
\num{2e-5}     &    0.1098       \\ 
\num{3e-5} & 0.1753\\
\num{4e-5} & 0.4222\\
\num{5e-5} & 0.0081 \\
\num{6e-5} & 0.0013\\
\num{7e-5} & 0.0022 \\
\num{8e-5} & 0.0008 \\
\num{9e-5} & 0.0008 \\
\num{1e-4} & 0.0026\\ \hline
\end{tabular}
\caption{$\mathbf{H}$ values calculated from simulation point patterns depicted in Figs. ~\ref{fig:fig4} and \ref{fig:fig5}} 
\label{table:DenSweepHVals}
\end{table}

\begin{table}[H]
\centering
\begin{tabular}{|c|c|c|c|}
\hline
System & nm/px & $\rho$ ( particles/$nm^{2}$) & Average particle radius (nm) \\ \hline
\romannumeral 1        &  11.76        &  0.000028        &   85.72       \\

\romannumeral 3       &  9.09        &  0.00004        &   71.78       \\
\romannumeral 4        &   11.76       &   0.00005       &  58.47        \\
\romannumeral 5        &    11.76      &  0.000048        &   73.8       \\
\romannumeral 6        &     11.76     &  0.000065        &   50.62       \\
\romannumeral 7        &  9.52        &  0.000096        &     40.58     \\

\romannumeral 8        &  9.52        &   0.000069       &    53.58      \\
\romannumeral 9        &     11.76     &  0.000067        &   53.79       \\
\romannumeral 10        &   12.5       &  0.000035        &    77.71      \\
\romannumeral 11        &  9.52        &  0.000134        &    35.25      \\
\romannumeral 12        &    9.52      &   0.000045       &    52      \\
 \hline
\end{tabular}
\caption{Parameters extracted from the TEM images in Ref~\cite{Prum2003jeb.00431}}
\label{table:TEMParameters}
\end{table}

\begin{figure}
    \centering
    \includegraphics[width=1\linewidth]{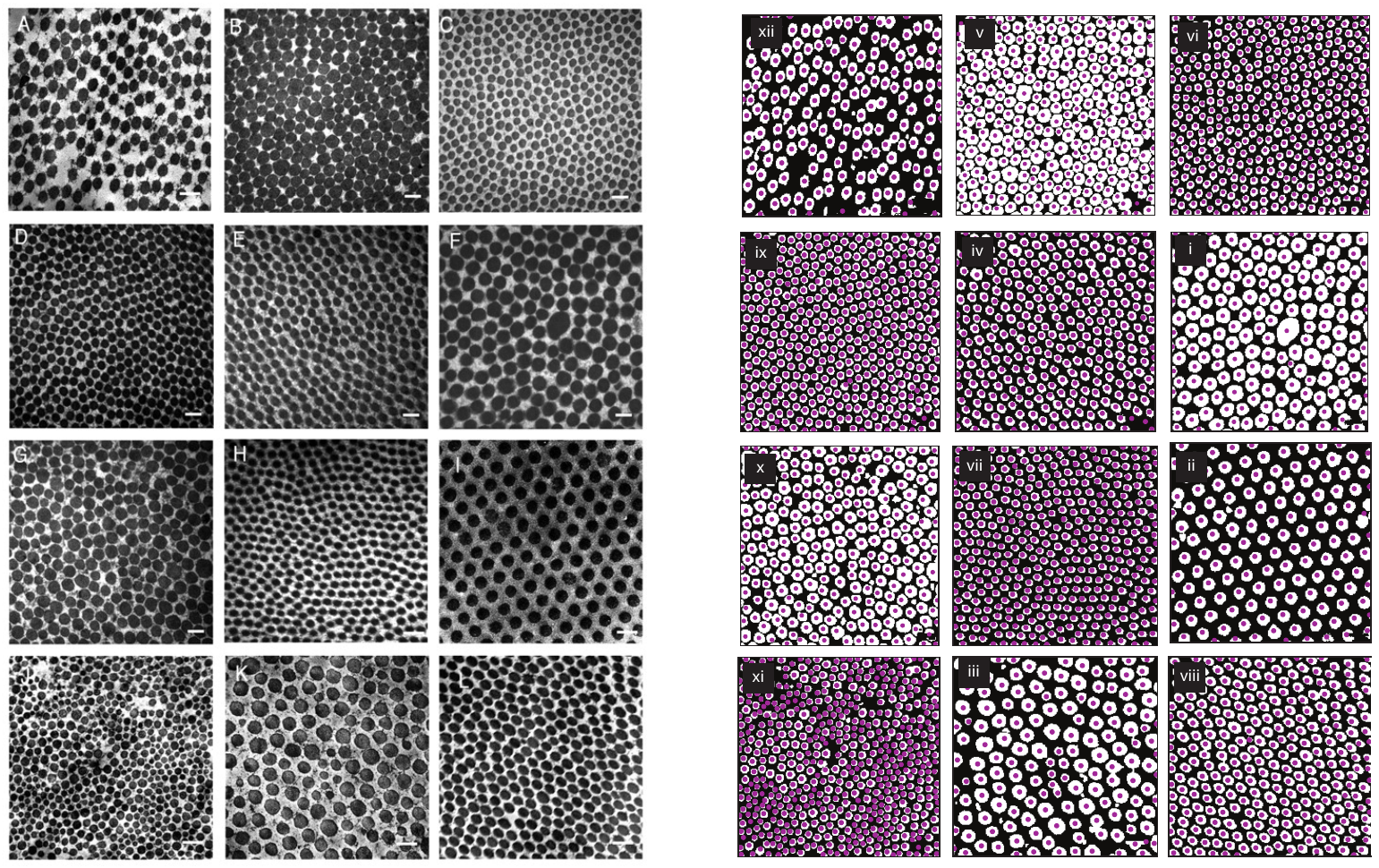}
    \caption{Left: reproduced with permission from Fig.~7 of Prum and Torres~\cite{Prum2003jeb.00431},
\textit{J.\ Exp.\ Biol.} \textbf{206}(14), 2409--2429 (2003),
\url{https://doi.org/10.1242/jeb.00431}.
Used with permission of The Company of Biologists Ltd, from ``Structural colouration of avian skin: convergent evolution of coherently scattering dermal collagen arrays,'' R.\ O.\ Prum and R.\ Torres, vol.\ 206, 2003; permission conveyed through Copyright Clearance Center, Inc. Right: same images, binarized with fibril centers located in magenta. Labeled using schema matching earlier figures.}
    \label{fig:sup1}
\end{figure}

\begin{figure}
    \centering
    \includegraphics[width=1\linewidth]{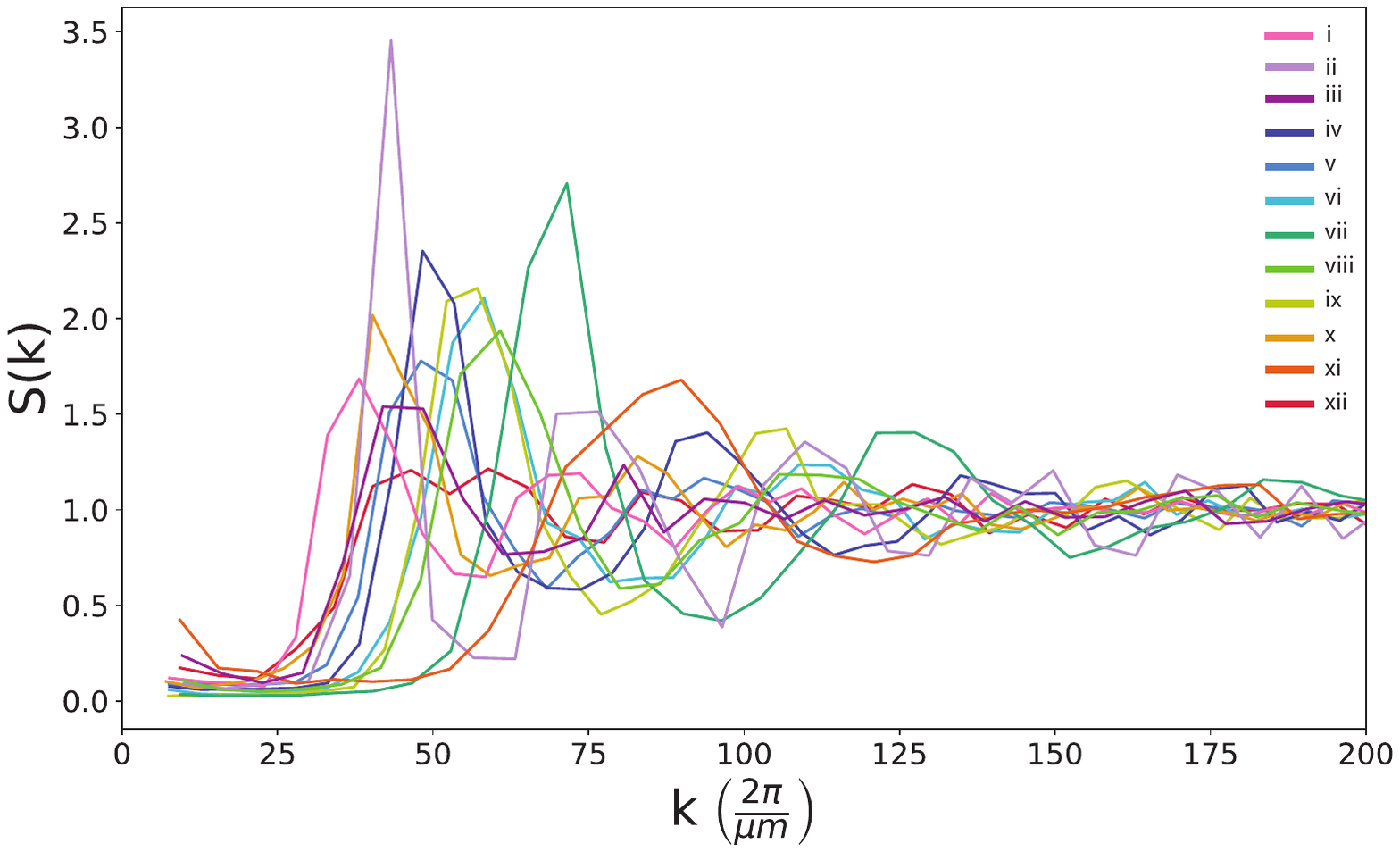}
    \caption{S(k) for TEM images in \ref{fig:sup1}}
    \label{fig:sup2}
\end{figure}

\foreach \p in {1,...,2}{%
  \begin{figure}[p]
    \centering
    \includegraphics[page=\p,width=\linewidth,height=0.85\textheight,keepaspectratio]{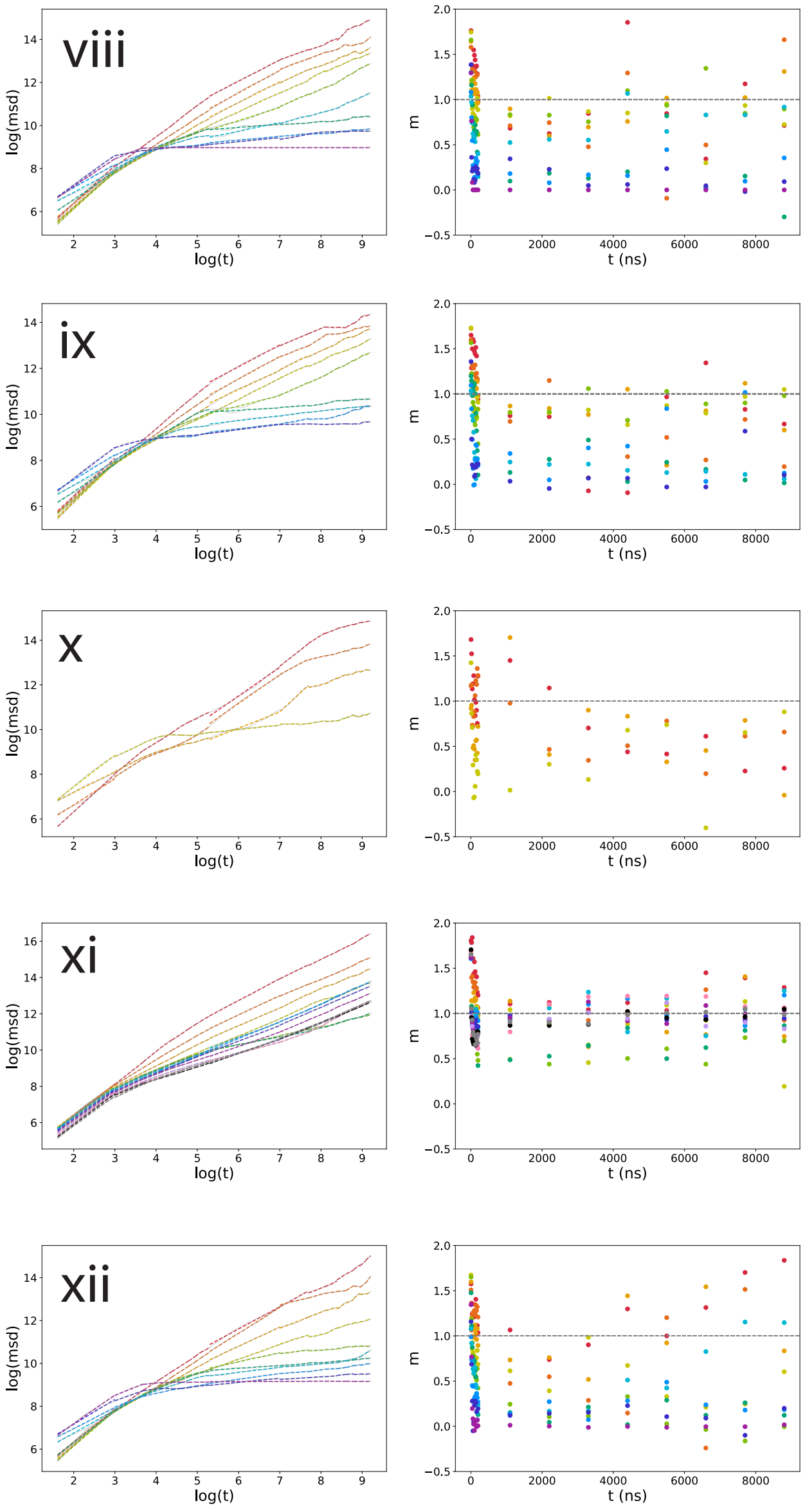}
    \ifnum\p=2
      \caption{Left column: loglog plots of mean squared displacements (MSD) for density sweep simulations. Right: slope of MSD. Brownian motion: $m=1$. Sub-diffusion: $m<1$. Super-diffusion: $m>1$.}
      \label{fig:sup3}
    \fi
  \end{figure}
}

\foreach \p in {1,...,5}{%
  \begin{figure}[p]
    \centering
    \includegraphics[page=\p,width=\linewidth,height=0.85\textheight,keepaspectratio]{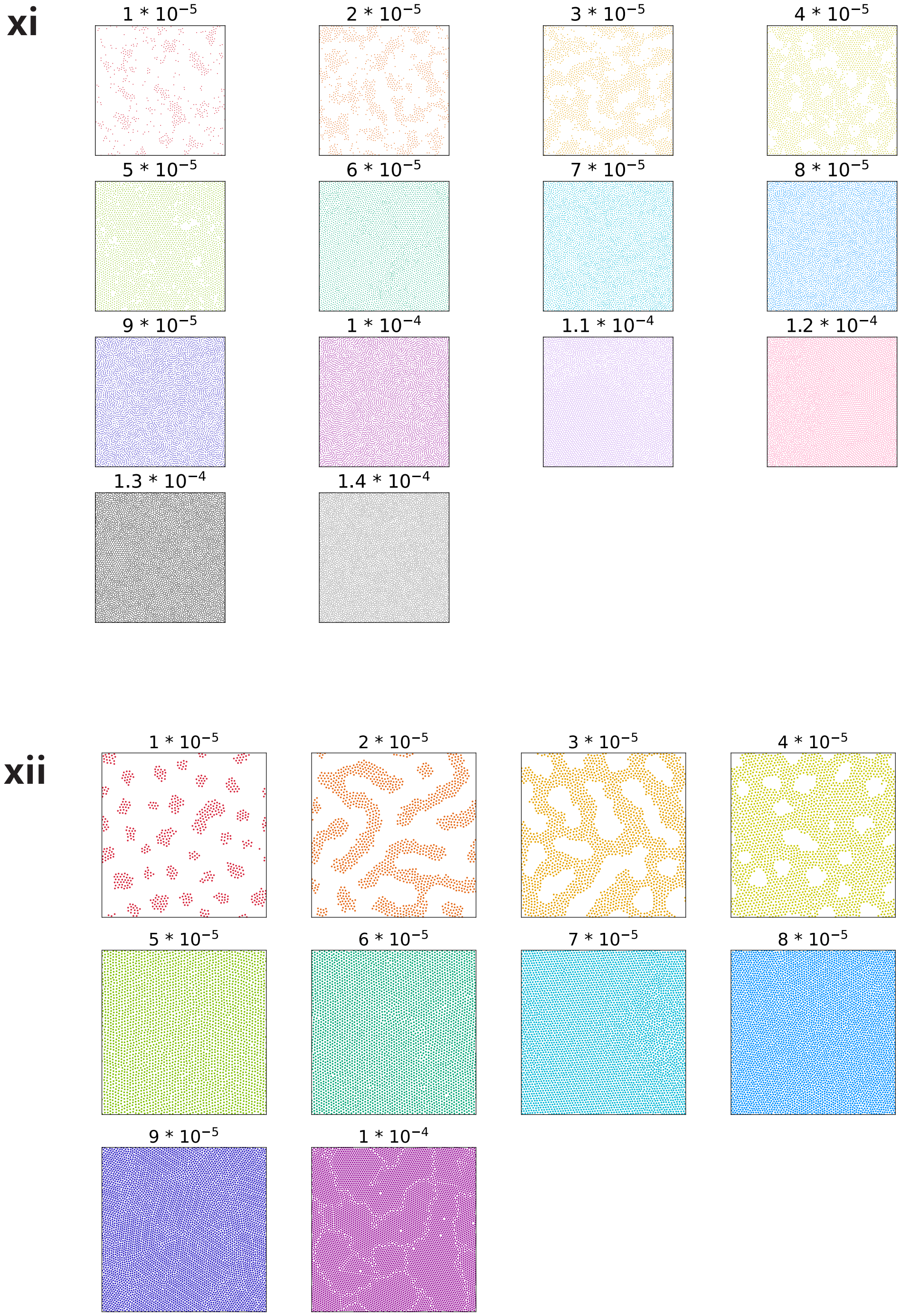}
    \ifnum\p=5
      \caption{hd+HNC simulations at 10 $\mu$s for systems i, iii--vii, ix--xii. ii was not modeled given its crystallinity, and viii is shown in Fig.~\ref{fig:fig4}.}
      \label{fig:sup4}
    \fi
  \end{figure}
}

\foreach \p in {1,...,6}{%
  \begin{figure}[p]
    \centering
    \includegraphics[page=\p,width=\linewidth,height=0.85\textheight,keepaspectratio]{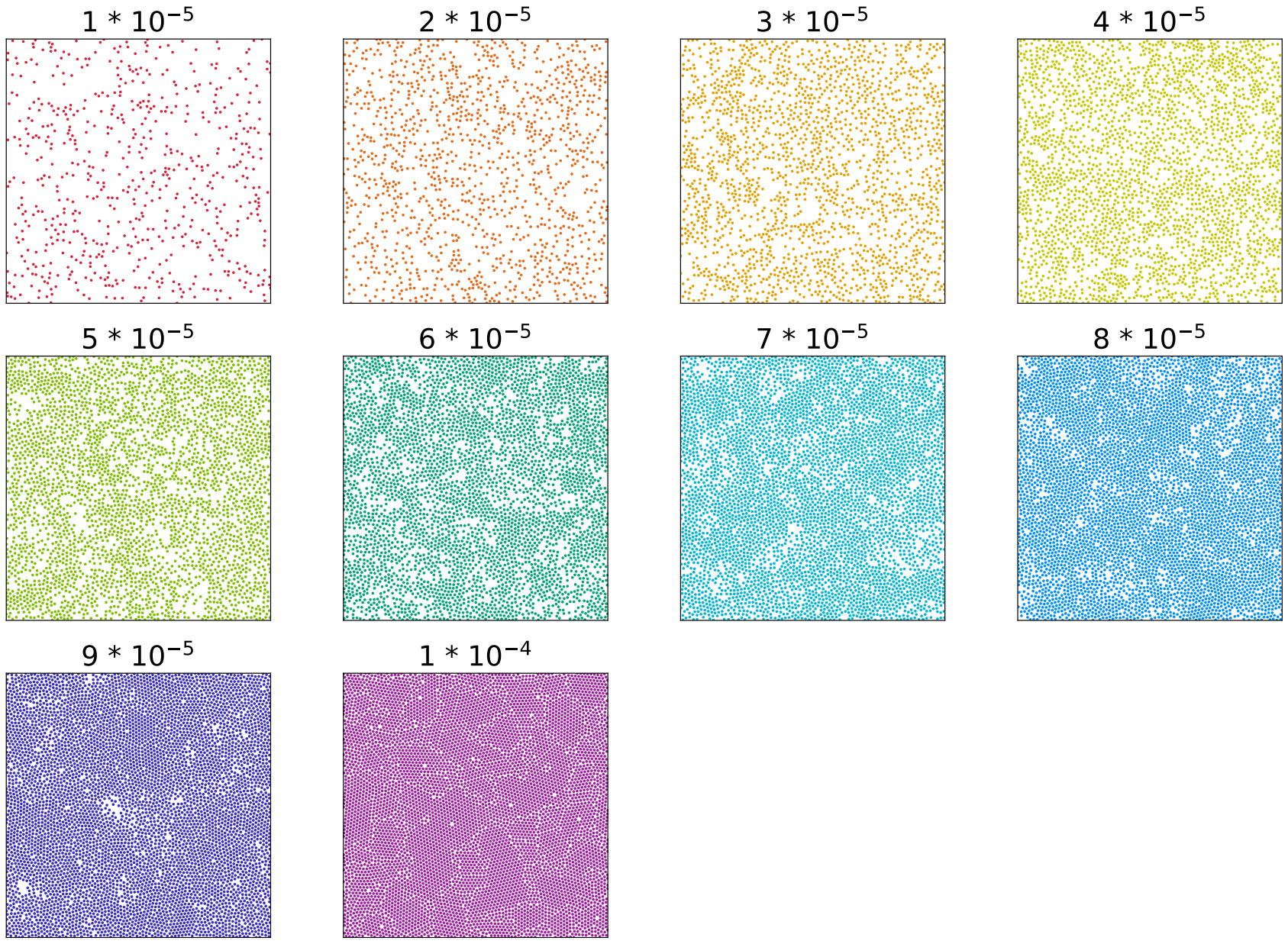}
    \ifnum\p=6
      \caption{hd simulations at 1 $\mu$s for systems i, iii--xii. }
      \label{fig:sup5}
    \fi
  \end{figure}
}

\begin{figure}
    \centering
    \includegraphics[width=1\linewidth]{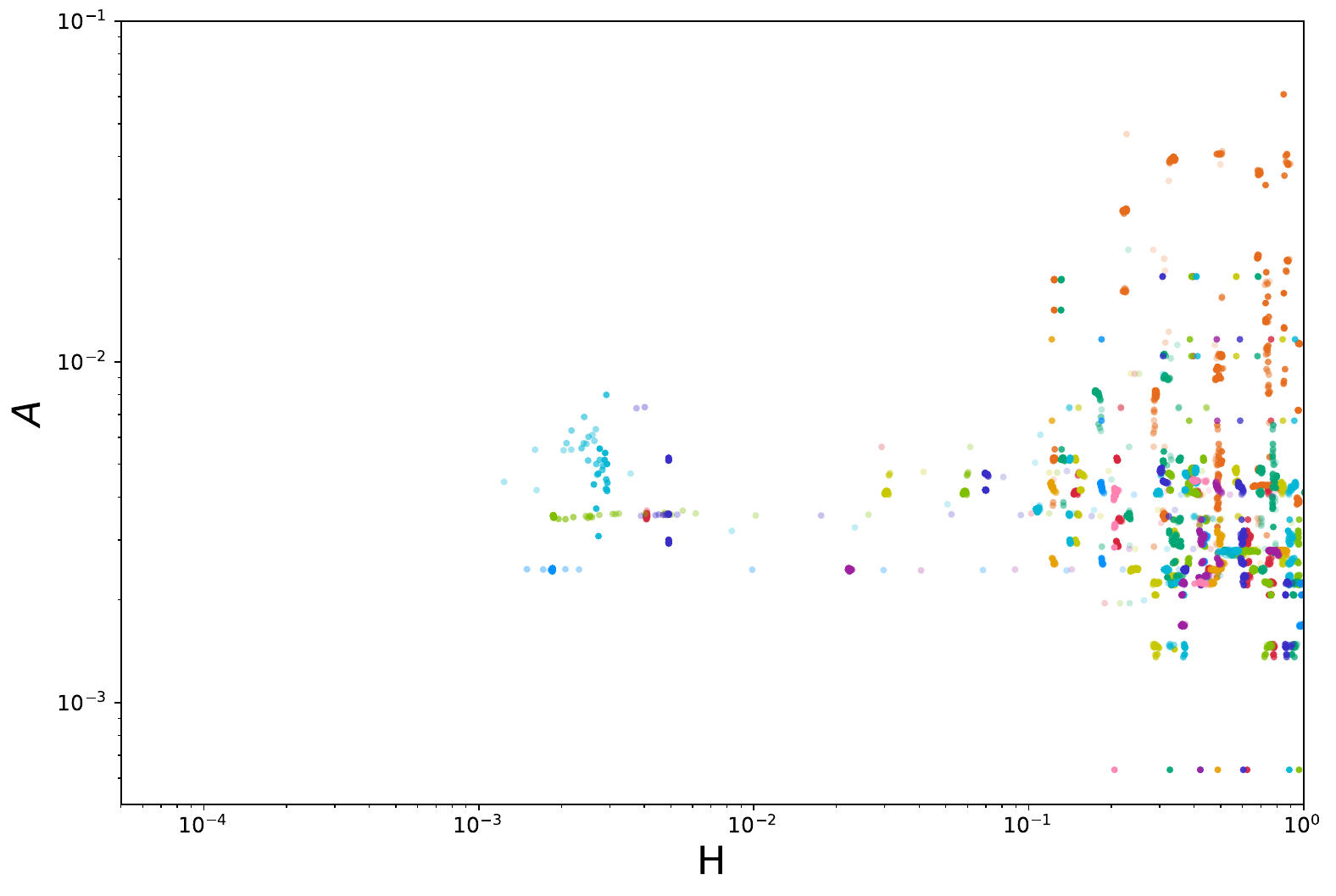}
    \caption{$A$ vs $\mathbf{H}$ for $\rho$ sweeps for systems with hd potentials.}
    \label{fig:sup6}
\end{figure}

\end{document}